\documentclass[12pt,apj]{emulateapj}

\begin{document}


\title{Evidence for Magnetic Flux Saturation in Rapidly Rotating M Stars}


\author{A. Reiners\altaffilmark{*}} 
\affil{Institut f\"ur Astrophysik, Georg-August-Universit\"at, D-37077
  G\"ottingen, Germany\email{Ansgar.Reiners@phys.uni-goettingen.de}}
\altaffiltext{*}{Emmy Noether Fellow}

\and

\author{G. Basri}
\affil{Astronomy Department, University of California, Berkeley, CA
  94720 \email{basri@berkeley.edu}}

\and

\author{M. Browning\altaffilmark{1}}
\affil{Astronomy Department, University of California, Berkeley, CA
  94720 \email{mbrowning@berkeley.edu}}
\altaffiltext{1}{Present address: Dept of Astronomy and Astrophysics,
  University of Chicago, Chicago, IL 60637}



\begin{abstract}
  We present magnetic flux measurements in seven rapidly rotating M
  dwarfs. Our sample stars have X-ray and H-alpha emission indicative
  of saturated emission, i.e., emission at a high level independent of
  rotation rate. Our measurements are made using near-infrared FeH
  molecular spectra observed with HIRES at Keck.  Because of their
  large convective overturn times, the rotation velocity of M stars
  with small Rossby numbers is relatively slow and does not hamper the
  measurement of Zeeman splitting. The Rossby numbers of our sample
  stars are as small as 0.01. All our sample stars exhibit magnetic
  flux of kilo-Gauss strength. We find that the magnetic flux
  saturates in the same regime as saturation of coronal and
  chromospheric emission, at a critical Rossby number of around 0.1.
  The filling factors of both field and emission are near unity by
  then. We conclude that the strength of surface magnetic fields
  remains independent of rotation rate below that; making the Rossby
  number yet smaller by a factor of ten has little effect. These
  saturated M-star dynamos generate an integrated magnetic flux of
  roughly 3 kG, with a scatter of about 1 kG. The relation between
  emission and flux also has substantial scatter.
\end{abstract}







\section{Introduction}

Stellar magnetic activity is closely related to rotation in solar-type
stars. In slow rotators, activity scales with the rotation rate until
it becomes saturated at a certain velocity, which means that it does
not grow further regardless of rotation rate \citep{Noyes84,
  Pizzolato03}. This effect is seen in a broad variety of activity
indicators \citep{Vilhu84, Vilhu87}. The threshold rate at which
saturation occurs depends on the spectral type of the star, with the
convective overturning time perhaps determining this threshold.
Saturation sets in where the value of the Rossby number $Ro =
P/\tau_{\rm conv} \la 0.1$, i.e.  where the timescale of rotation is
significantly shorter than the timescale of typical convective eddies.

Indicators of stellar activity are usually coronal or chromospheric
emission observed at X-ray, UV, optical, infrared, or radio
wavelengths. We know from the Sun that this emission is induced by
magnetic fields heating the upper layers of the solar atmosphere, and
by analogy we conclude that stellar activity is connected to magnetic
fields on the surface of other stars. At high rotation rates (or small
Rossby numbers), all activity indicators saturate, i.e. they do not
grow over a certain level regardless of higher rotation rates
\citep[see, e.g.,][]{James00}. Two possible explanations exist for the
saturation: 1) The stellar dynamo process saturates and stars cannot
produce magnetic fields stronger than the saturation value; or 2) The
magnetic fields continue to grow at more rapid rotation, but the
fraction of the surface filled with fields -- or the area covered by
spots -- reaches unity so that no more emitting plasma can be placed
on the star. The only way to decide which way the stars go is to
directly measure the magnetic field.  Unfortunately, this is very
difficult and the picture -- particularly in stars with saturated
activity -- is not yet clear.

\citet{Saar96} has collected measurements of magnetic flux $Bf$, of
the filling factor $f$, and of rotation periods to investigate the
behavior of magnetic fields on stars. It is important to realize that
the magnetic flux, or the magnetic field average over the whole
surface, is not the same as the local field. It is the (unsigned) mean
average of the magnetic field strength over the whole surface.
Furthermore, the strongest magnetic fields in cool spots may not be
fully captured because their contribution to the total flux is
dimished due to their low temperature.  As on the Sun, magnetic flux
is probably concentrated in relatively small regions of strong fields
\citep[see, e.g.][]{JKV00}. \citet{Saar96} shows that in stars
rotating slower than the saturation threshold the magnetic flux $Bf$
as well as the filling factor $f$ show the same trend as all other
activity indicators: They grow with larger rotation rate. At high
rotation rates, \citet{Saar96} claims that saturation occurs in the
filling factor $f$ but not in the magnetic flux $Bf$. From this result
one would conclude that a star rotating at the saturation threshold is
completely covered with magnetism ($f=1$), and that the saturation
phenomenon is due to the saturation of the emission process while the
star's magnetic flux can grow further with higher rotation.
\citet{Saar01} has reinvestigated this issue with a few more data
points, noting that there is some indication for a saturation at $Bf
\sim 3$\,kG at small Rossby numbers. Much higher magnetic flux could
be in contradiction to the idea that magnetic fields in stellar
atmospheres cannot grow stronger than the equipartition field, i.e.
the field strength at which the magnetic pressure equals the gas
pressure. However, \citet{Solanki94} shows that the equipartition
field may not necessarily be a hard upper limit for the field strength
at $f=1$, so that more rapidly rotating stars could in principle have
much stronger fields.

The only way to decide whether $Bf$ does saturate or not is to provide
direct measurements of magnetic flux in the regime of saturated
activity.  Generally, the measurement of magnetic fields relies on the
splitting of spectral lines through the Zeeman effect
\citep[e.g.][]{Robinson80}. In rapid rotators, the subtle effect of
magnetic broadening is buried under the rotational line broadening so
that it is particularly difficult to directly measure the magnetic
flux in the regime of saturated magnetic activity. However, the
crucial datum for rotational line broadening is the projected rotation
velocity $v\,\sin{i}$ and not the Rossby number or the rotation period
(which are the numbers that appear to set the saturation threshold of
magnetic activity).  Mainly because of the smaller radius of cooler
stars, the surface rotation velocity at which saturation sets in
depends on spectral type. In early G-type stars, the saturation
velocity in the transition region is on the order of 30\,km\,s$^{-1}$
\citep[and only 15\,km\,s$^{-1}$ in the corona,][]{Ayres99}. In M
dwarfs, it is less than 5\,km\,s$^{-1}$ \citep[see e.g.][]{Reiners07}.
Thus, in sun-like stars, the high surface rotation velocity required
for activity saturation hampers the measurement of magnetic flux, but
this regime can easily be probed in M stars.

One potential problem with the use of M dwarfs for the investigation
of dynamo related phenomena is that the interior structure changes
around spectral type M3.5 -- stars cooler than that are completely
convective. Nevertheless, no change in activity is observed at the
threshold to complete convection. A rotation-activity connection is
observed in M stars down to spectral types M~8.5 \citep{Mohanty03,
  Reiners07}.

In this paper, we present direct measurements of magnetic flux in
several M stars. Some of them are very rapid rotators and clearly
belong to the regime of saturated activity. We aim to clarify whether
the magnetic flux $Bf$ saturates as H$\alpha$ and X-ray emission do,
or if $Bf$ continues to grow beyond $Bf \sim 3$\,kG.

\section{Sample and Observations}

For our sample we chose a number of mid-M stars with known X-ray
emission and presumably high rotation rates. The values $v\,\sin{i}$
and $\log{L_{\rm X}/L_{\rm bol}}$ are taken from \citet{Delfosse98}
except for GJ\,3379. This value is calculated from the X-ray
luminosity taken from \cite{NEXXUS}. With one exception, projected
rotation velocities $v\,\sin{i}$ were available for all targets. We
chose only stars in which $v\,\sin{i}$ was reported to be above
5\,km\,s$^{-1}$ and which show saturated normalized X-ray emission.
For GJ\,3379 we are not aware of any former $v\,\sin{i}$ measurement,
but the high value of normalized X-ray emission is indicative of
saturation and we added the star to our sample.

\begin{deluxetable}{lccc}
  \tablecaption{\label{tab:observations} Log of observations.}
  \tablewidth{0pt}
  \tablehead{\colhead{Name} & \colhead{UTC Date} & \colhead{Exp.Time [s]} & \colhead{$\log{\frac{L_{\rm X}}{L_{\rm bol}}}$} }
  \startdata

  GJ 3379            & 2007-09-30 &  200 & $-2.86$\\
  GJ 2069\,B         & 2008-01-24 &  600 & $-2.77$\\
  Gl 493.1           & 2007-04-25 &  600 & $-3.31$\\
  LHS 3376           & 2007-04-25 & 1800 & $-3.63$\\
  GJ 1154\,A         & 2007-04-25 &  600 & $-3.28$\\
  GJ 1156            & 2007-04-25 &  600 & $-3.39$\\
  Gl 412\,B          & 2007-04-25 & 1200 & $-3.28$

  \enddata
\end{deluxetable}

Data were taken at the W.M. Keck observatory with the HIRES
spectrograph. Our setup covers the wavelength range from below
H$\alpha$ (6560\,\AA) up to the molecular absorption band of FeH
around 1\,$\mu$m. We used a slit width of 1.15\,\arcsec achieving a
resolving power of about $R = 31\,000$. Our log of observations is
given in Table\,\ref{tab:observations}. Data were cosmic-ray
corrected, flatfielded, background subtracted, and wavelength
calibrated using a ThAr spectrum. Data reduction was carried out using
routines from the \texttt{echelle} package within the ESO/MIDAS
distribution.  Fringing is not an issue in spectra taken with the new
HIRES CCD, even in very red spectral regions around 1\,$\mu$m.

\section{Analysis}
\label{sect:results}

\begin{deluxetable}{lcrrrr}
  \tablecaption{\label{tab:results} Results of our analysis. Literature values are from \citet{Delfosse98}.}
  \tablewidth{0pt}
  \tablehead{\colhead{Name} & \colhead{SpT} & \colhead{$v\,\sin{i}$} & \colhead{$Bf$} & \colhead{$\log{\frac{L_{{\rm H}\alpha}}{L_{\rm bol}}}$} & $v\,\sin{i}_{\rm Lit}$\\
  & & [km\,s$^{-1}$] & [G] & & [km\,s$^{-1}$]}
  \startdata

  GJ 3379            & M3.5 & $<3$ & 2300    & $-3.35$ &     \\
  GJ 2069\,B         & M4.0 & $ 6$ & 2700    & $-3.28$ &  9.2\\
  Gl 493.1           & M4.5 & $18$ & 2100    & $-3.46$ & 16.8\\
  LHS 3376           & M4.5 & $19$ & 2000    & $-3.73$ & 14.6\\
  GJ 1154\,A         & M5.0 & $ 6$ & 2100    & $-3.55$ &  5.2\\
  GJ 1156            & M5.0 & $17$ & 2100    & $-3.53$ &  6.5\\
  Gl 412\,B          & M6.0 & $ 5$ & $>$3900 & $-3.72$ &  7.7
  \enddata
\end{deluxetable}

The analysis of our spectra follows the strategy laid out in
\citet{RB06} and Reiners \& Basri (2007, RB07 in the following). We
measure the equivalent width of the H$\alpha$ emission and convert
this number to normalized H$\alpha$ luminosity using M-star
atmospheres calculated with the PHOENIX code \citep{Allard01}. To
measure the projected rotation velocity $v\,\sin{i}$ and the magnetic
flux $Bf$ of our sample stars, we utilize the absorption band of
molecular FeH close to 1\,$\mu$m. We compare our data to spectra of
the slowly rotating M-stars GJ\,1002 (M5.5) and
Gl\,873\footnote{Gl\,873 is rotating at $v\,\sin{i} \sim
  3$\,km\,$^{-1}$ (RB07), not at a higher velocity as reported in
  \cite{Delfosse98}. The small (but detectable) rotation of Gl~873
  does not affect our measurements.} (M3.5). In order to match the
absorption strength of the target spectra, the intensity of the FeH
absorption lines in the two comparison spectra is modified according
to an optical-depth scaling \citep[see][]{RB06}. In a first step, we
compare the artificially broadenend spectrum of GJ\,1002 to the target
spectra in the wavelength region at 9930--9960\,\AA\ to determine the
value of $v\,\sin{i}$ by $\chi^2$-minimization.

For the determination of the magnetic flux $Bf$, we concentrate on
smaller wavelength regions that contain absorption lines particularly
useful for this purpose, i.e. regions that contain some magnetically
sensitive as well as magnetically insensitive lines. The magnetic flux
of Gl\,873 was measured to be 3.9\,kG \citep[using an atomic FeI
line;][]{JKV00}. For our measurement, we are using a spectrum that
contains both the FeI line and the FeH absorption band. The FeI line
in this spectrum is consistent with the same magnetic flux value as
found by \cite{JKV00} so that we can use the FeH pattern for the
calibration of magnetic flux measurements in other stars. This method
does not require theoretical models of the magnetic Zeeman splitting
of FeH lines, which are not available yet. 

\begin{figure}
  \centering
  \includegraphics[width=.48\hsize]{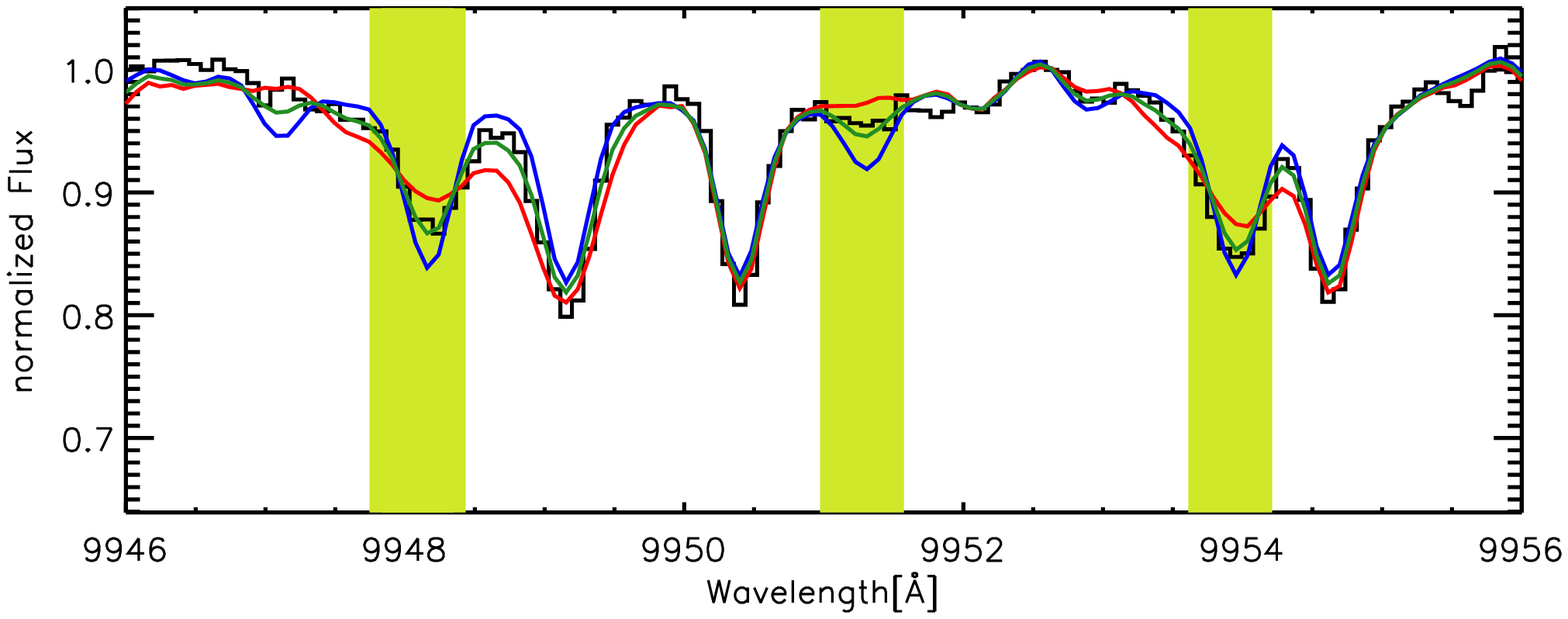}
  \includegraphics[width=.48\hsize]{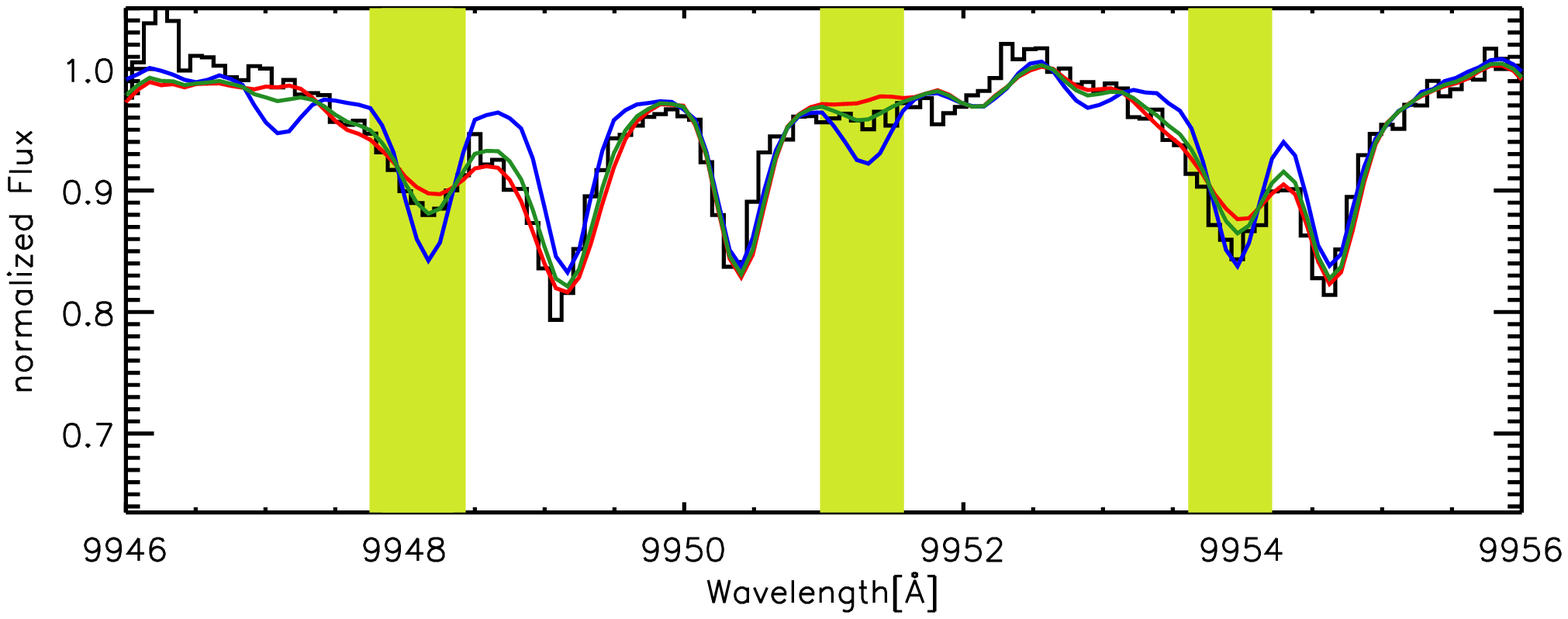}\\
  \includegraphics[width=.475\hsize,clip=,bbllx=20,bblly=0,bburx=600,bbury=425]{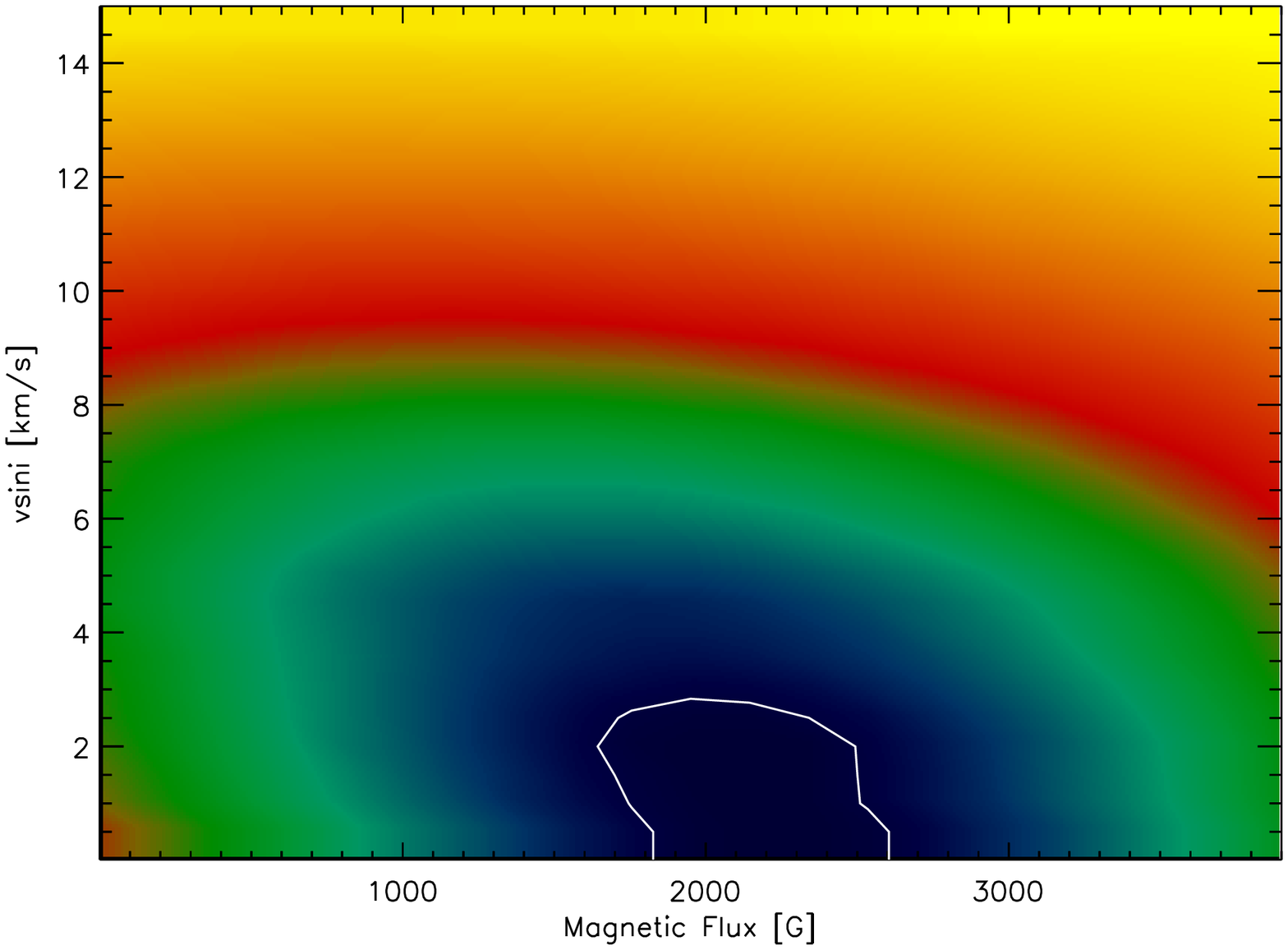}
  \includegraphics[width=.475\hsize,clip=,bbllx=20,bblly=0,bburx=600,bbury=425]{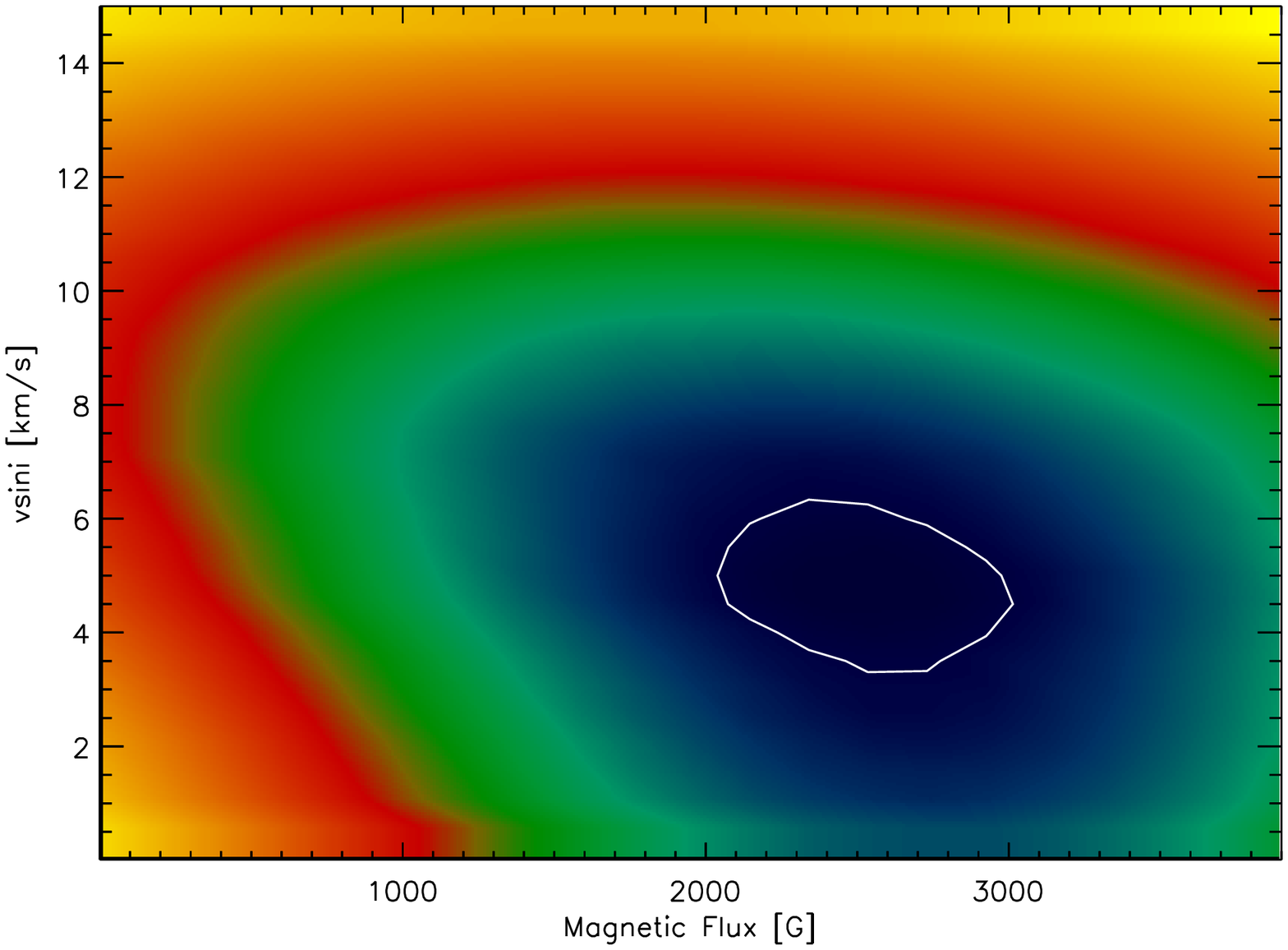}
  \caption{\label{fig:slow}Top panel: Data and fit of the two slow
    rotators GJ 3379 (left) and GJ 2069B (right). The data are shown
    in black. Our fit to the data for the case of no magnetic flux is
    overplotted in blue, very strong magnetic flux ($Bf = 3.9\,kG$) in
    red, and the best fit with intermediate flux values in green.
    Bottom panel: $\chi^2$-landscapes showing the goodness of fit as a
    function of $v\,\sin{i}$ and $Bf$. Dark/blue color indicates good
    fit quality, bright/yellow color means bad fit. The white contour
    marks the formal 3$\,\sigma$ region, i.e., $\chi^2 < \chi^2_{\rm
      min} + 9$.}
\end{figure}

\begin{figure}
  \centering
  \includegraphics[width=.48\hsize]{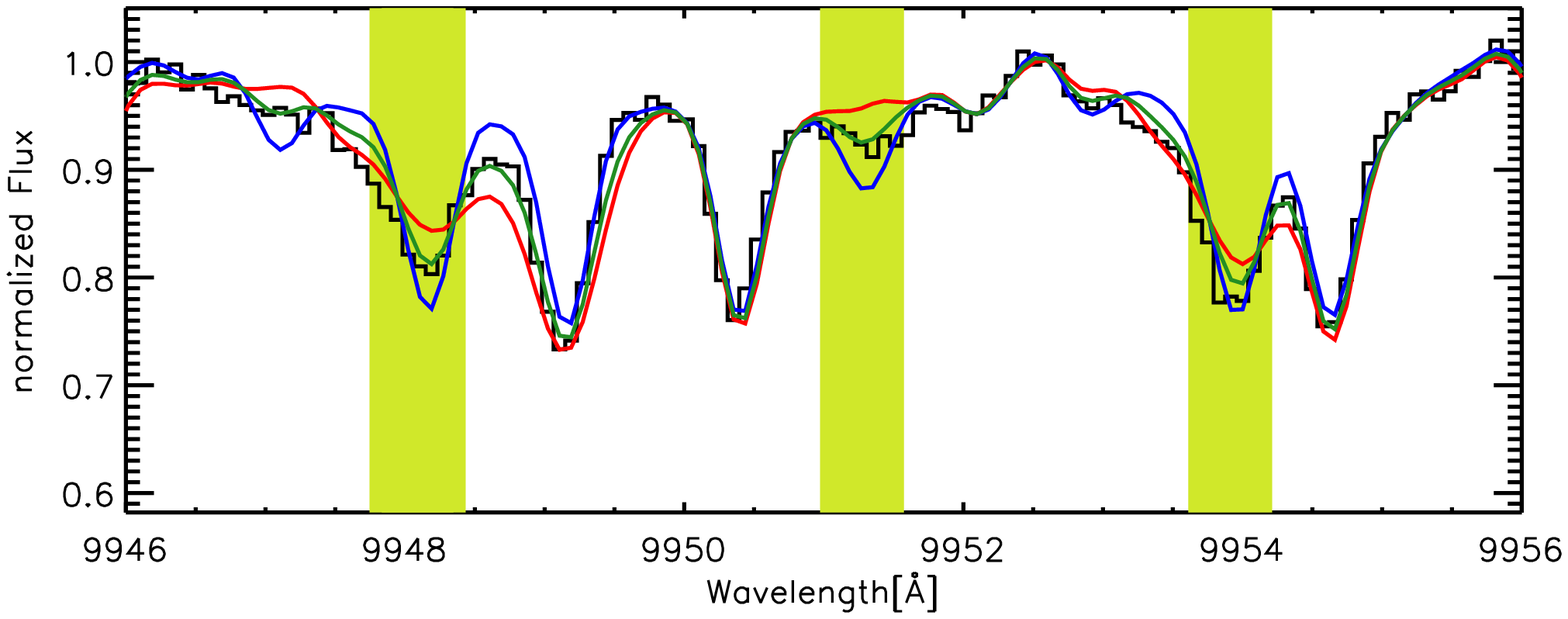}
  \includegraphics[width=.48\hsize]{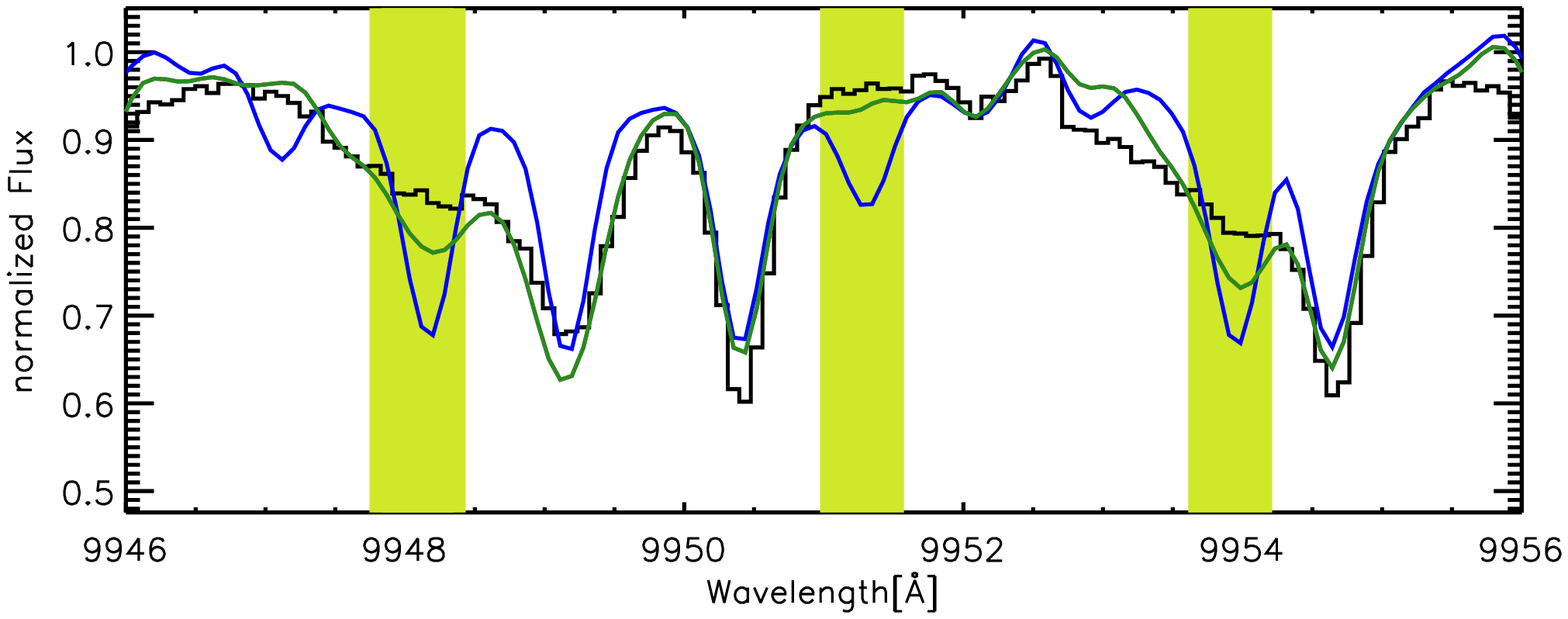}\\
  \includegraphics[width=.475\hsize,clip=,bbllx=20,bblly=0,bburx=600,bbury=425]{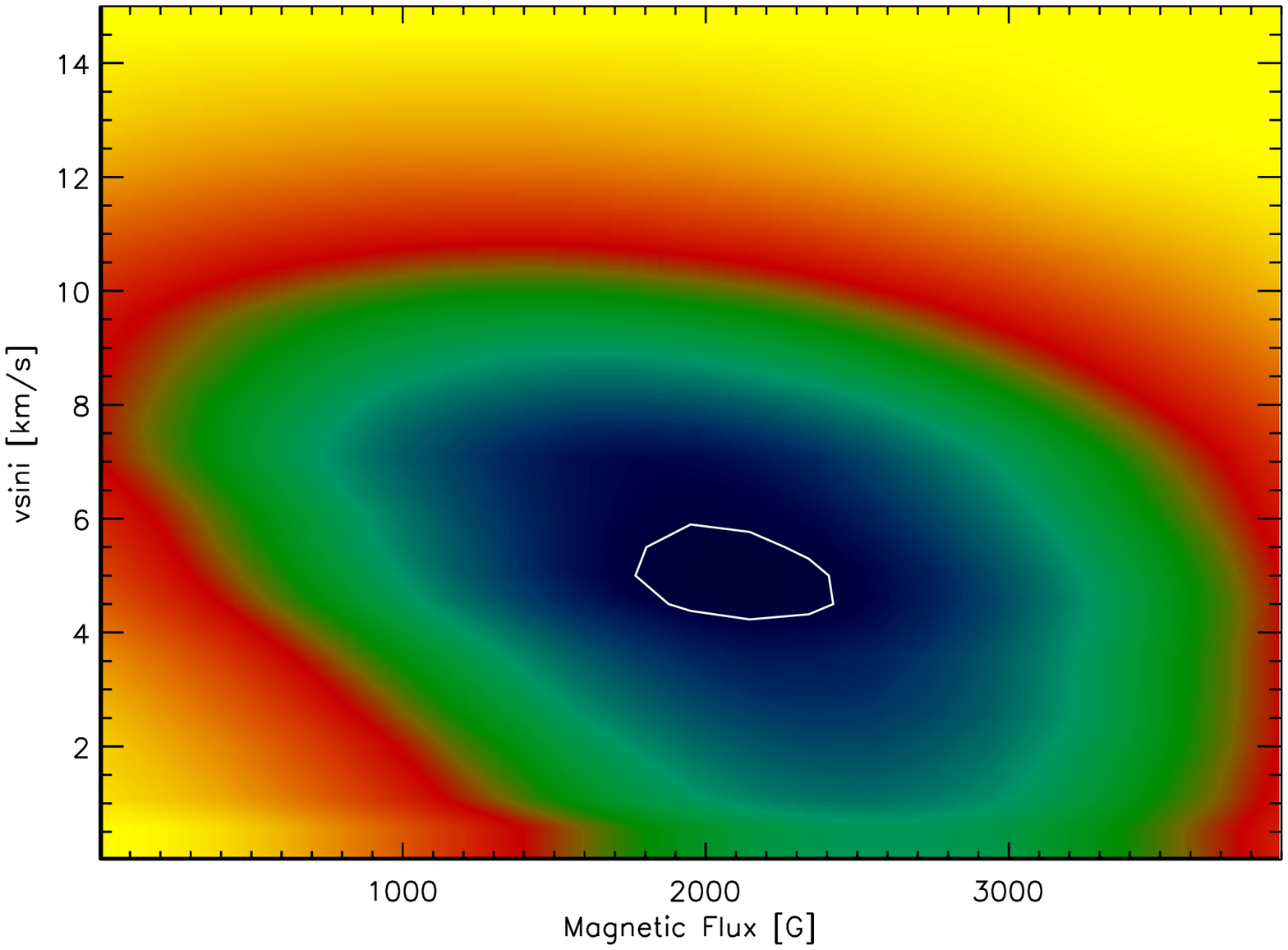}
  \includegraphics[width=.475\hsize,clip=,bbllx=20,bblly=0,bburx=600,bbury=425]{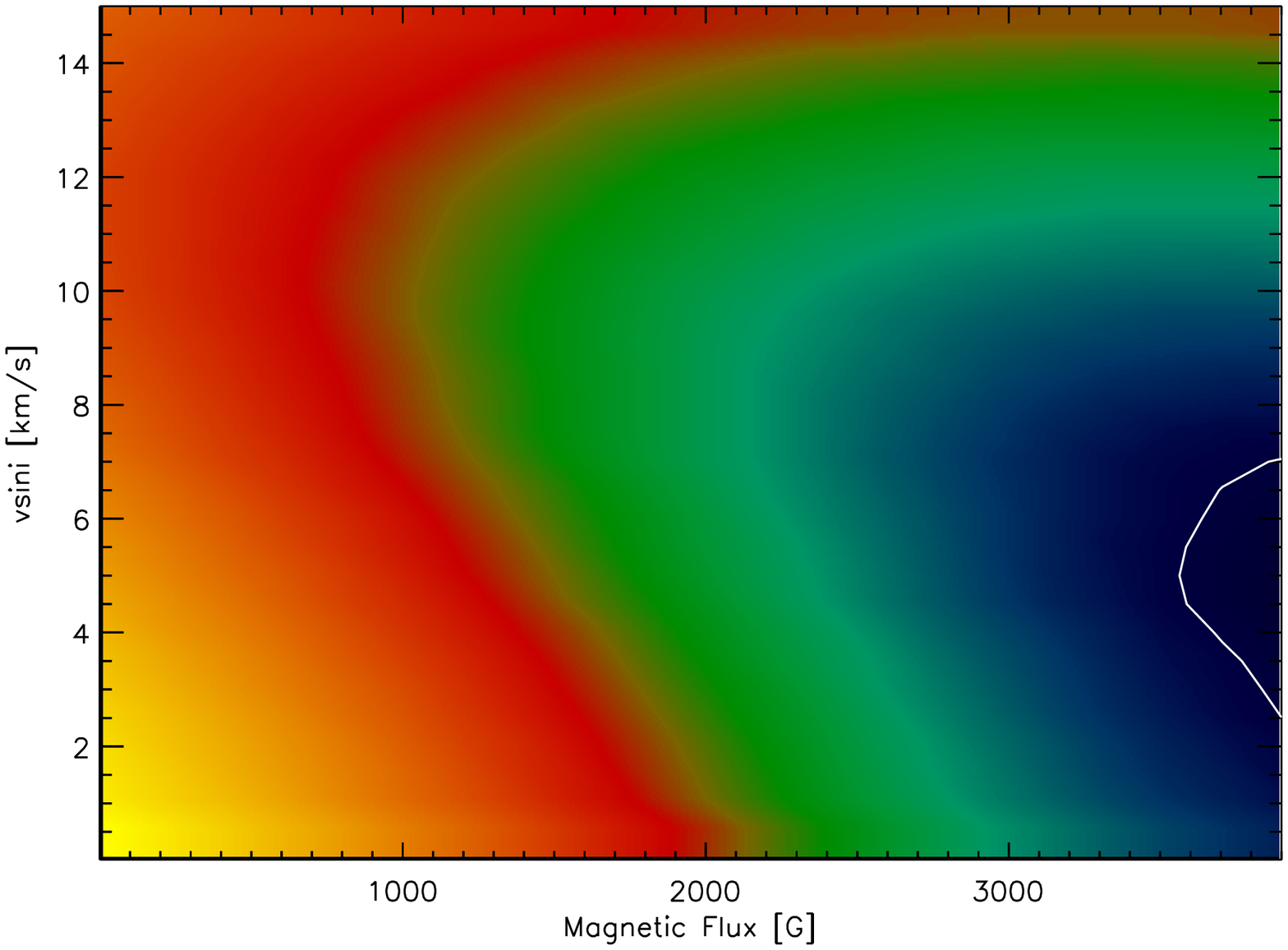}
  \caption{\label{fig:intermediate}Date and fit (top panel) and
    $\chi^2$-landscapes (bottom panel) as in Fig.\,\ref{fig:slow} for
    the two intermediate rotators GJ 1154A (left) and GJ 412B
    (right).}
\end{figure}

We determine the magnetic flux of our target stars by comparison of
the spectral regions at 9895.5--9905.5\,\AA, 9937.5--9941.0\,\AA,
9946.0--9956.0\,\AA, and 9971.5--9981.0\,\AA\ (for more details see
RB07). In Figs.\,\ref{fig:slow}--\ref{fig:rapid2} we show the data and
the quality of our fit in the top panels.  Note that in rapid rotators
the difference between magnetic and non-magnetic stars is not
necessarily clearest at the exact location of magnetically sensitive
lines. The blending of lines through rotation pronounces differences
at wavelengths where the equivalent widths of lines differ the most
between active and inactive stars.

For example, the two FeH lines at 9949.1\,\AA\ and 9951.7\,\AA\ are
magnetically not very sensitive \citep[see][]{RB06, Reiners08}, but it
appears that the 9949.1\AA\ line effectively gains a little in
equivalent width. Thus, at rotation rates as high as shown in
Figs.\,\ref{fig:rapid} and \ref{fig:rapid2}, the region around
9949\,\AA\ becomes the one of largest difference. The reason for this
are the magnetically sensitive features next to insensitive lines
together with the effective gain in equivalent width, although such
differences are not necessarily overt for the observer once Doppler
broadening is introduced. On the other hand, it can also happen that
at wavelengths where the unrotated spectra are quite different,
rotational broadening averages in adjacent flux in such a way that the
magnetic differences end up erasing each other there (e.g. at
9948\,\AA).  Thus, it is necessary to carry out a spectral fitting
procedure after proper preparation instead of relying on appearances
in original template spectra at fixed wavelenghts.

\section{Results}

\begin{figure}
  \centering
  \includegraphics[width=.46\hsize]{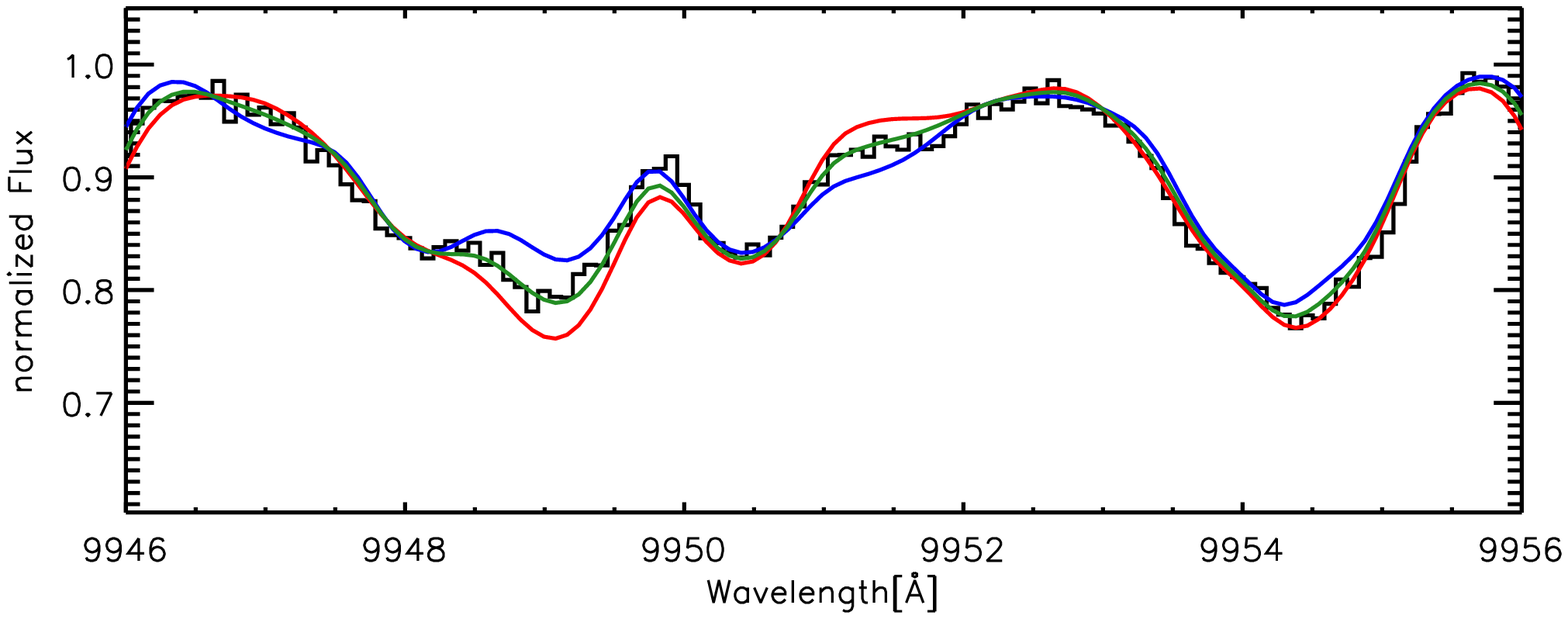}\quad\,
  \includegraphics[width=.46\hsize]{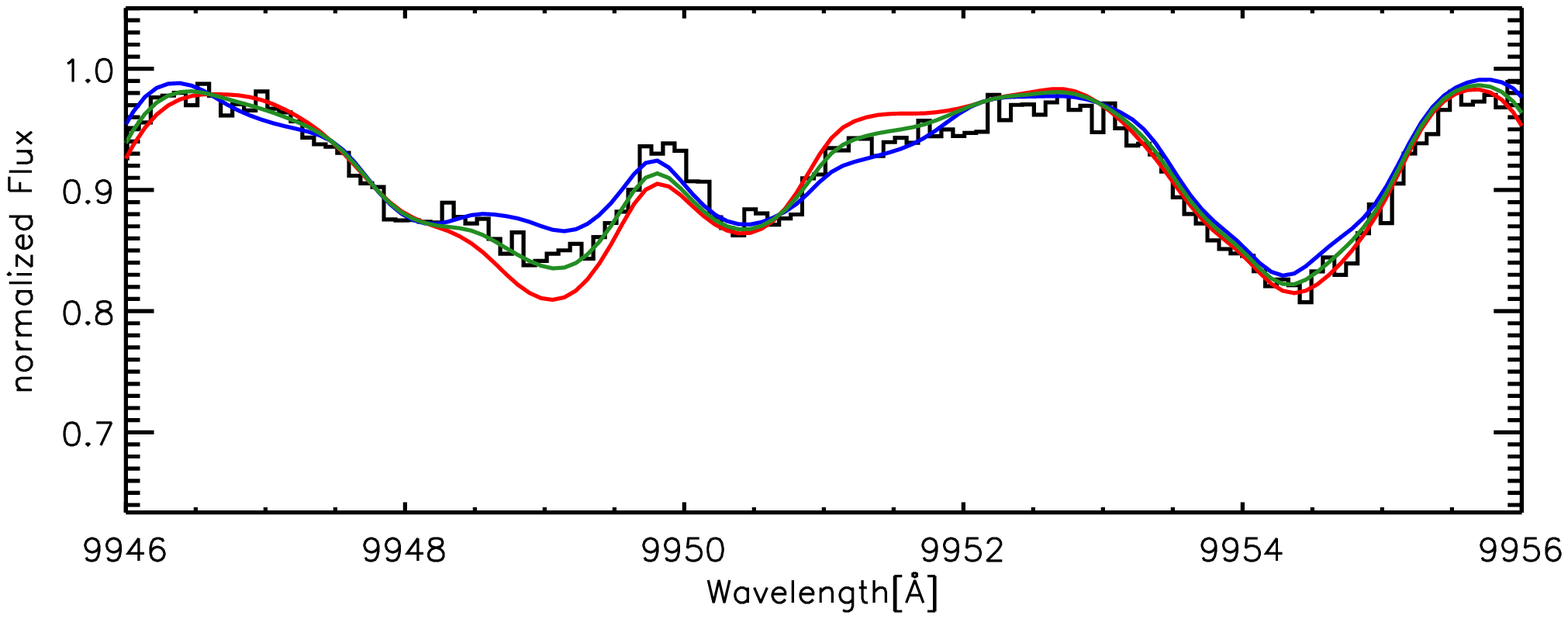}\\
  \includegraphics[width=.475\hsize,clip=,bbllx=20,bblly=0,bburx=600,bbury=425]{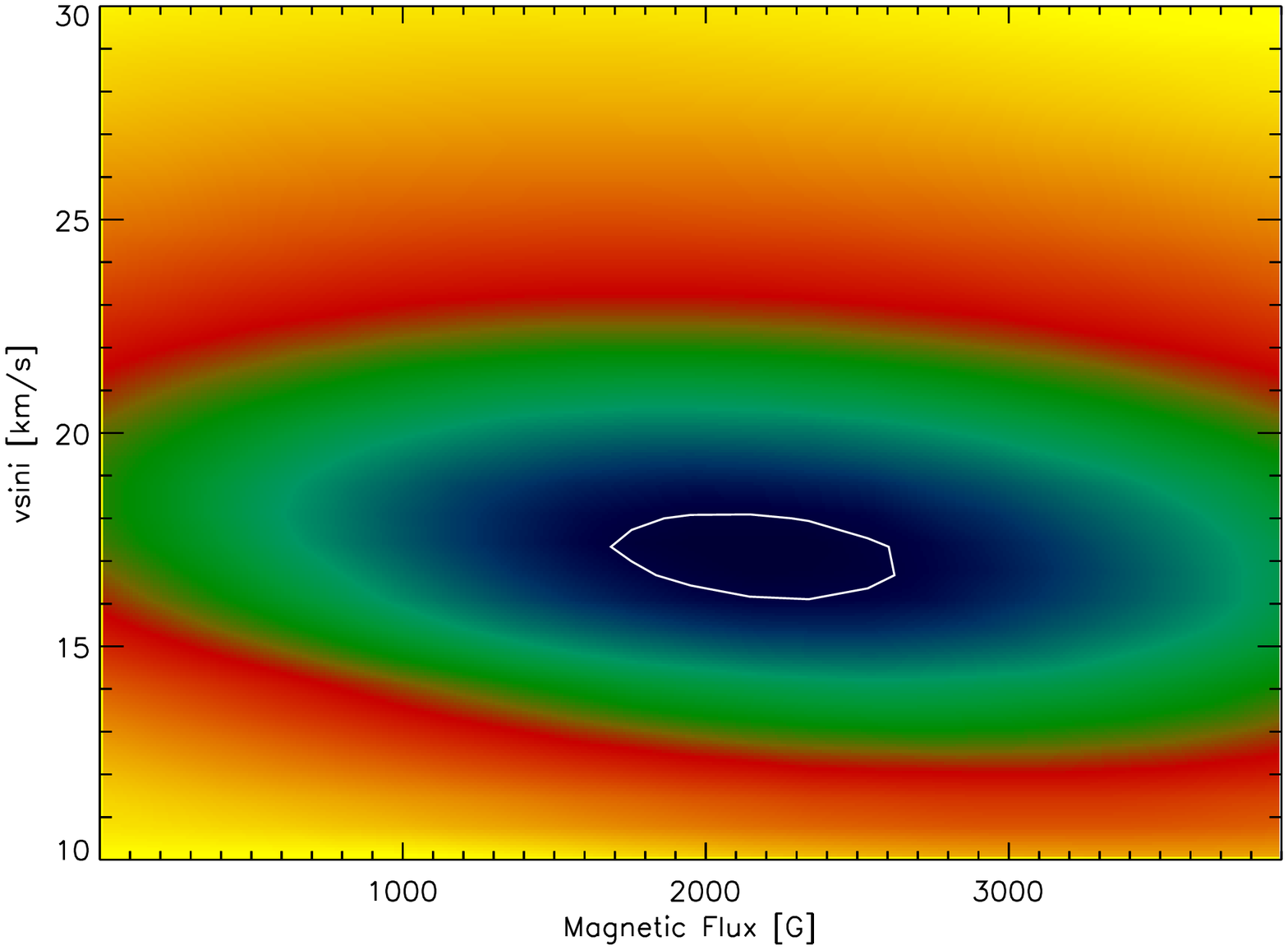}
  \includegraphics[width=.475\hsize,clip=,bbllx=20,bblly=0,bburx=600,bbury=425]{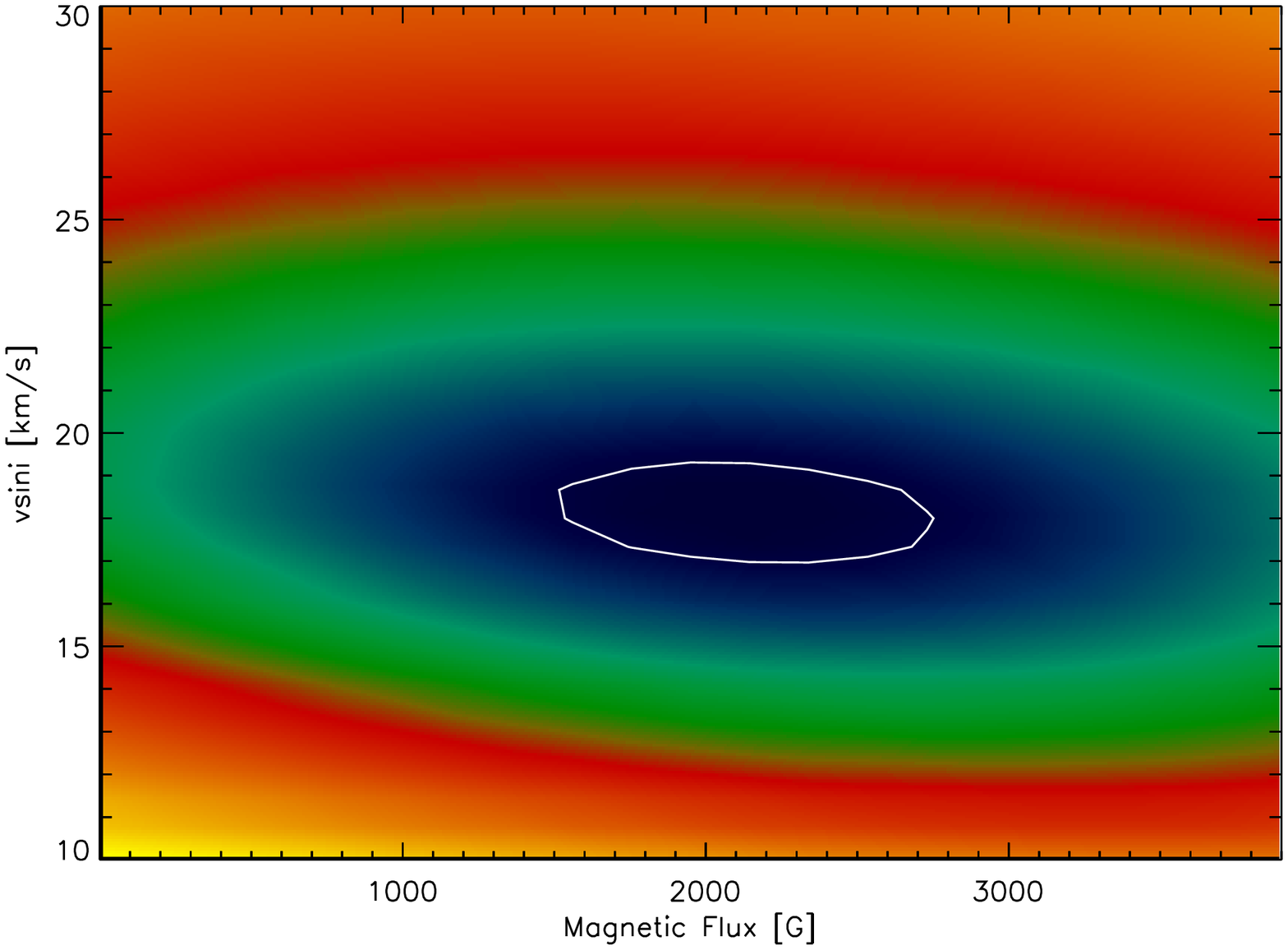}
  \caption{\label{fig:rapid}Data and fits (top panel) and
    $\chi^2$-landscape as in Fig.\,\ref{fig:slow} for the rapid
    rotators GJ 1156 (left) and Gl 493.1 (right).}
\end{figure}

\begin{figure}
  \includegraphics[width=.46\hsize]{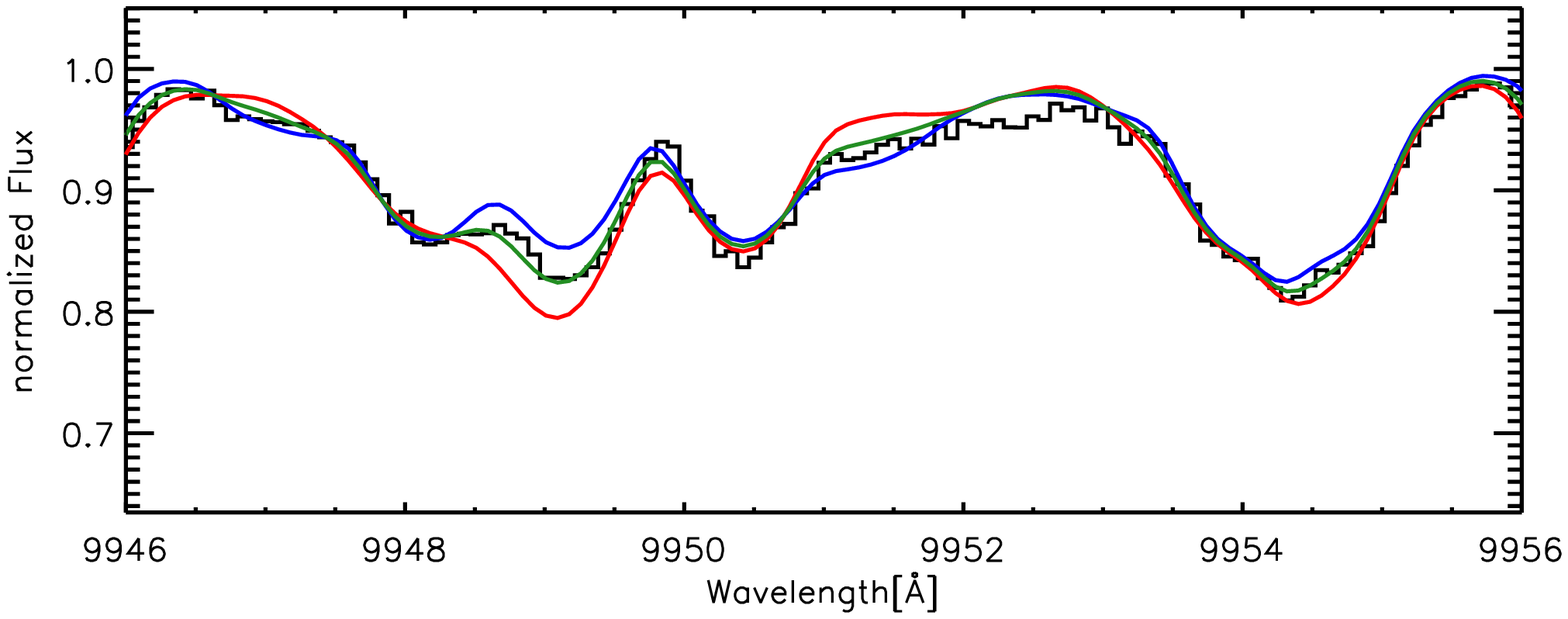}\\
  \includegraphics[width=.475\hsize,clip=,bbllx=20,bblly=0,bburx=600,bbury=425]{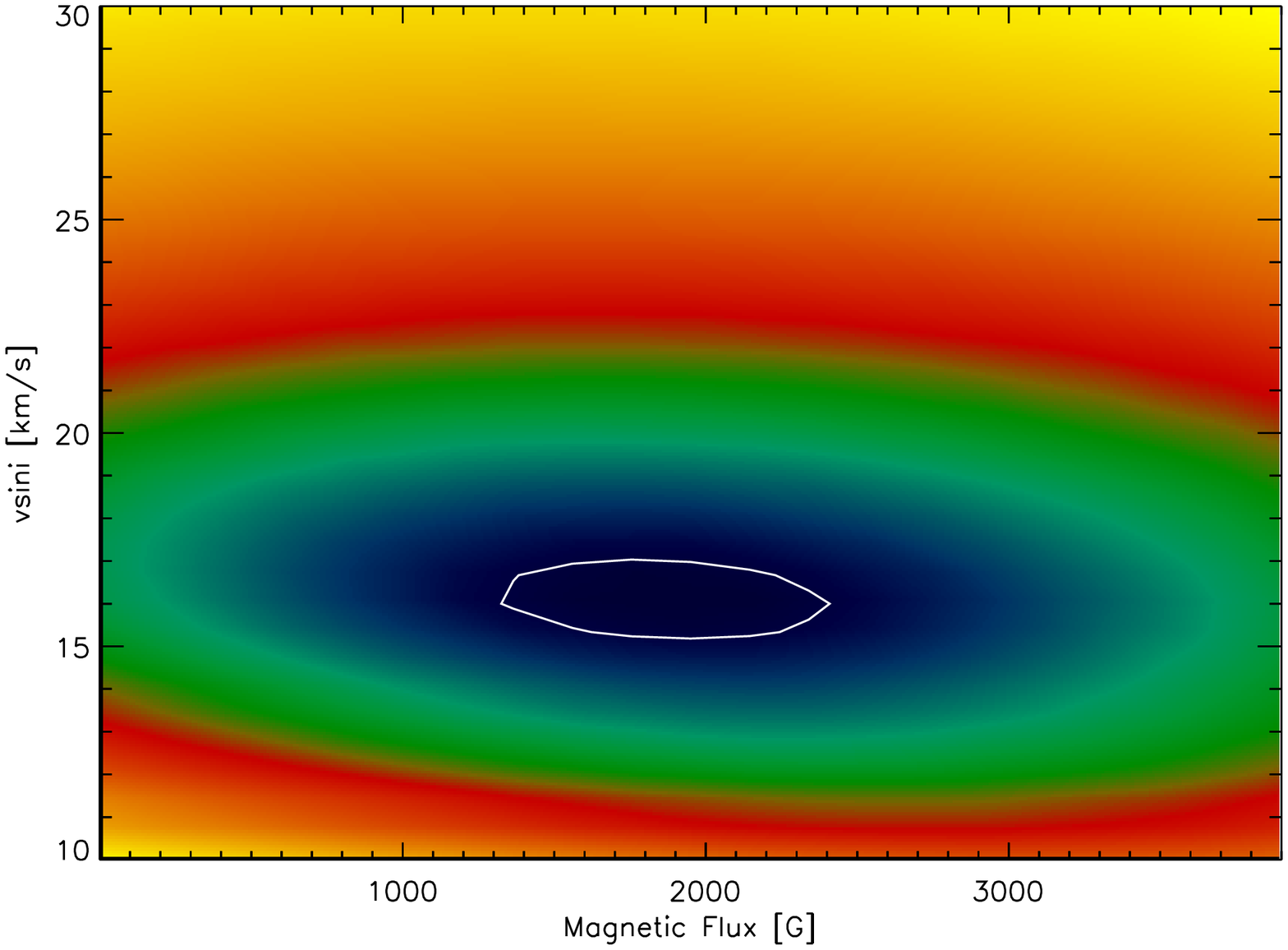}
  \caption{\label{fig:rapid2}As Fig.\,\ref{fig:rapid} for the rapid
    rotator LHS 3376.}
\end{figure}

In the bottom panels of Figs.\,\ref{fig:slow}--\ref{fig:rapid2}, we
show the $\chi^2$-landscapes for all our targets as a function of
$v\,\sin{i}$ and $Bf$. Color-coding displays the quality of the fit.
In each $\chi^2$-landscape, the white contour marks the 3$\sigma$
region, i.e. $\chi^2 < \chi^2_{\rm min} + 9$ within this region. In
all cases, the mean deviation per degree of freedom is on the order of
1 ($\chi_{\nu} \approx 1$) for the estimated signal-to-noise ratio.
Uncertainties in $v\,\sin{i}$ and $Bf$ are typically around
1\,km\,s$^{-1}$ and a few hundred Gauss, respectively. We emphasize
that in particular in the case of $Bf$ systematic errors are a more
severe source of uncertainty so that the total uncertainty in $Bf$ is
more realistically in the 500--1000\,G range.

The results of our analysis are given in Table\,\ref{tab:results};
projected rotation velocity $v\,\sin{i}$, magnetic flux or the mean
magnetic field $Bf$, and normalized H$\alpha$ activity $\log{L_{{\rm
      H}\alpha}/L_{\rm bol}}$ are given in columns 3, 4, and 5,
respectively. All H$\alpha$ measurements confirm that our sample
targets are active stars close to the activity saturation level. Three
stars show very rapid rotation on the order of 20\,km\,s$^{-1}$, three
stars are rotating at a velocity around 6\,km\,s$^{-1}$. GJ~3376 shows
rotation below our detection limit of $v\,\sin{i} \approx
3$\,km\,s$^{-1}$.  For comparison, we include in column 6 measurements
of $v\,\sin{i}$ by \cite{Delfosse98}.  In GJ\,1156, we measure a
rotational velocity three times higher than formerly reported, and in
GJ~2069\,B our new value of rotational broadening is $v\,\sin{i} =
6$\,km\,s$^{-1}$ while \citet{Delfosse98} reports $v\,\sin{i} =
9$\,km\,s$^{-1}$. We attribute the differences to higher data quality
in our sample and to our more sophisticated fitting procedure. The
uncertainty in $v\,\sin{i}$ also depends on the magnetic flux with a
higher uncertainty at higher magnetic flux. The crosstalk between
magnetic flux and rotation velocity may be weaker in other wavelength
regions (but so far this was not investigated in detail).

Measuring magnetic flux is hampered if a target is rapidly rotating
because resolving individual lines becomes difficult in the presence
of rotation. In the three rapid rotators Gl~493.1, LHS~3376, and
GJ~1156, we used only the spectral regions 9946--9956\,\AA\ and
9972--9981\,\AA, which are particularly well suited in rapid rotators
\citep{RB06}. 

In all seven stars, we detected mean magnetic fields of 2\,kG or
stronger. RB07 investigated 22 M dwarfs, 17 of them have spectral
types earlier than M7.  All 8 of them with normalized H$\alpha$
luminosity larger than $\log{L_{{\rm H}\alpha}/L_{\rm bol}} = -4$ also
have mean magnetic fields on the order of 2\,kG or more. The most
active stars in that sample exhibit significant rotation, but the
sample contains only one star rotating more rapidly than $v\,\sin{i} =
10$\,km\,s$^{-1}$.  Our results are in good agreement with the
relation between mean magnetic fields and normalized H$\alpha$
activity found in RB07. The (projected) rotation velocities of the
three most rapidly rotating stars in our sample are at least a factor
of two higher, but we see no sign of normalized H$\alpha$ luminosity
higher than in RB07, and no exceptionally high value of $Bf$.
\emph{The main result of this work is that none of the three rapid
  rotators with projected rotation velocities close to
  20\,km\,s$^{-1}$ shows a mean magnetic field above 3\,kG.}

Can we really measure magnetic fields in stars rotating as rapidly as
$v\,\sin{i} = 20$\,km\,s$^{-1}$? In \citet{RB06} we have shown that
the magnetic sensitivity of FeH in principle allows the measurement of
magnetic flux in stars rotating as rapidly as $v\,\sin{i} =
30$\,km\,s$^{-1}$. The limiting factor in rapid rotators is the
achievable signal-to-noise ratio. In our case of ``only'' $v\,\sin{i}
\approx 20$\,km\,s$^{-1}$ this is not a crucial problem. The
differences between stars with strong and weak magnetic fields in the
presence of rapid rotation can be seen in the spectra we show in
Figs.\,\ref{fig:rapid} and \ref{fig:rapid2}. The results of our
$\chi^2$ fits plotted in the lower panels of Fig.\,\ref{fig:rapid} and
\ref{fig:rapid2} show that in all three stars $\chi^2$ becomes
significantly larger at very low field strengths or if one allows for
mean fields as high as 4\,kG.

Our active template star Gl~873, which we are using as comparison for
our targets, has a mean magnetic field of about 4\,kG. From comparison
to the spectrum of Gl\,873 we cannot measure magnetic flux in excess
of that value. In RB07 we show a spectrum of YZ~CMi (M4.5, $v\,\sin{i}
= 5$\,km\,s$^{-1}$) from which we inferred a mean field stronger than
4\,kG.  This spectrum shows that in the presence of stronger fields
the magnetically sensitive lines can become even more washed out
following the principles of Zeeman broadening.\footnote{We show this
  spectrum in Fig.\,3 of RB07.  In that plot, the ratio of the
  magnetically sensitive to the insensitive lines is smaller in YZ~CMi
  than in our magnetically active template.} The spectrum of Gl~412\,B
(WX~UMa) shows the same behavior at a rotation velocity of $v\,\sin{i}
= 5$\,km\,s$^{-1}$.  Thus, although we lack a spectrum independently
calibrated to stronger magnetic flux to compare with, we see no way
that the spectra of the three most rapidly rotating stars can be
consistent with magnetic flux stronger than $Bf = 4$\,kG. We discuss
these stars further below.

\subsection{Saturation of the magnetic flux $Bf$}

\begin{figure}
  \plottwo{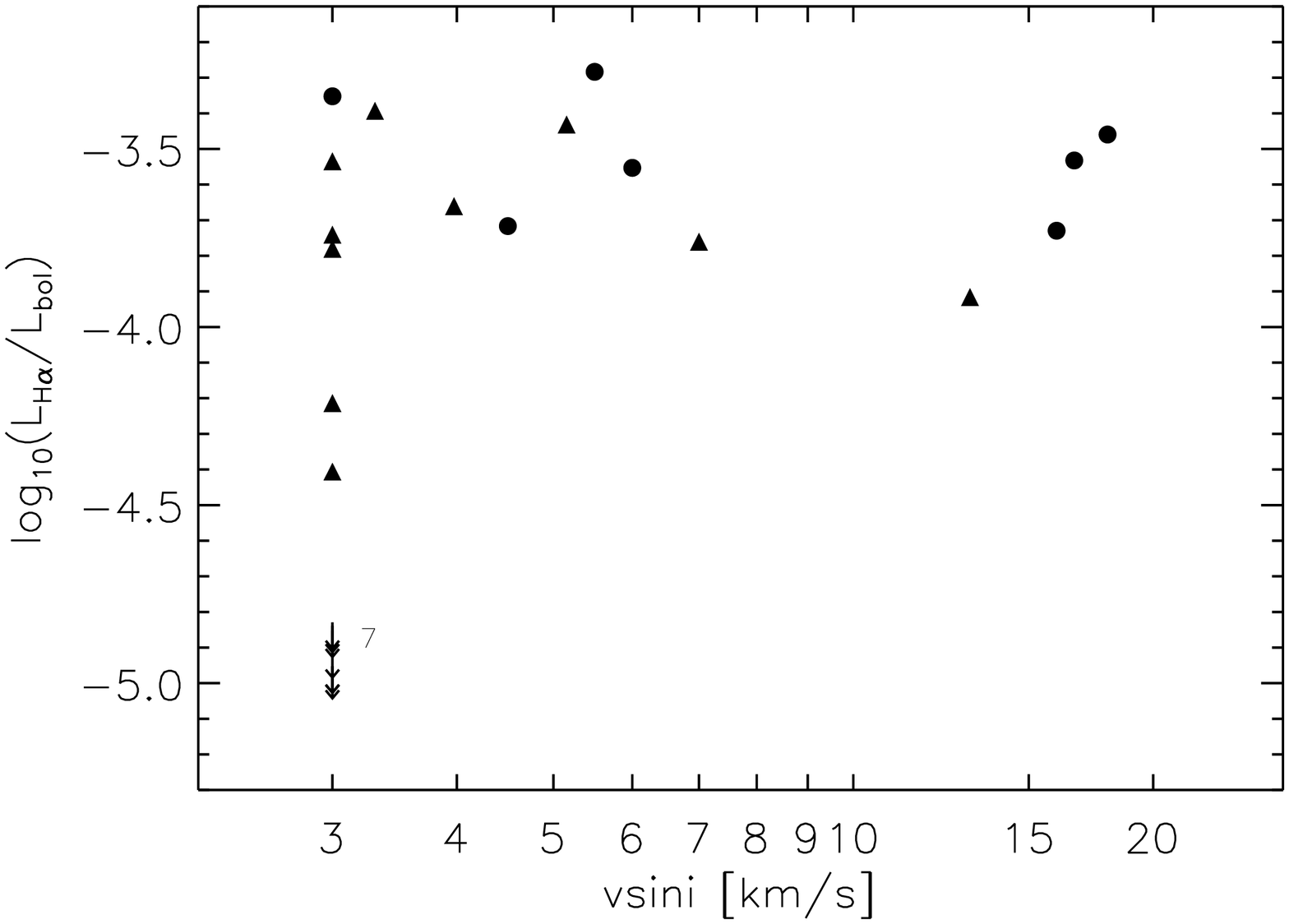}{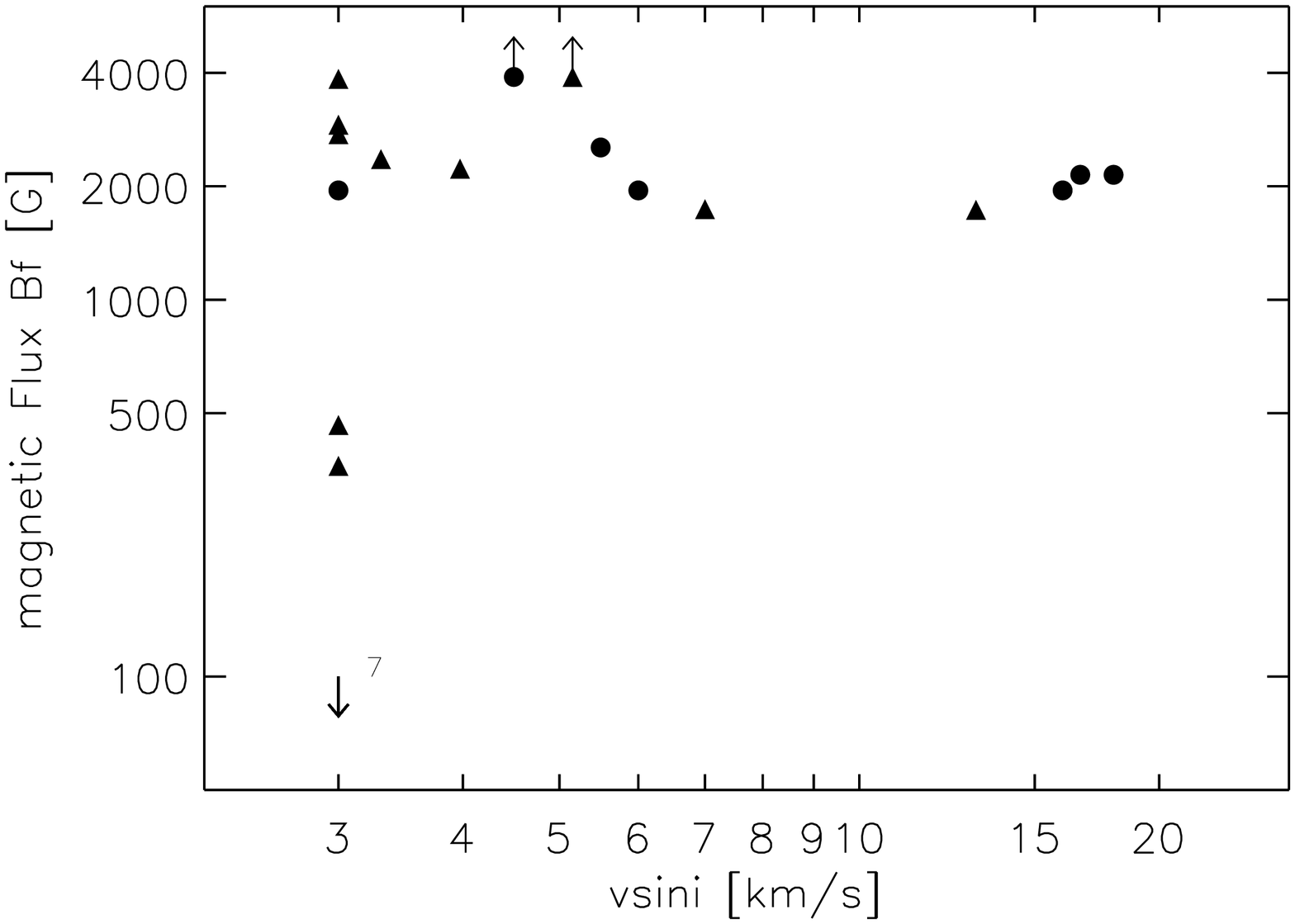}
  \caption{\label{fig:vsini_alpha} Left panel: Normalized H$\alpha$
    activity as a function of $v\,\sin{i}$ in M stars; right panel:
    Magnetic flux $Bf$ as a function of $v\,\sin{i}$. Filled circles
    are from this work, triangles come from RB07. Downward arrows
    indicate upper limits, numbers give the number of multiple
    measurements with the same results. The two lower limits of $Bf$
    are indicated with upward arrows. }
\end{figure}

It is well accepted that coronal and chromospheric activity saturate
at high rotation rates. \citet{Mohanty03} showed that this relation is
still valid in stars as late as spectral type M8.5. In mid-M stars,
all stars with detected rotational broadening show H$\alpha$ emission
at the saturation level. In the left panel of
Fig.\,\ref{fig:vsini_alpha}, we plot $v\,\sin{i}$ vs.  $\log{L_{{\rm
      H}\alpha}/L_{\rm bol}}$ for all stars of spectral types earlier
than M7 from RB07 together with the stars from our new sample.  This
shows that the rotation-activity relation is still valid in the
combined sample. In the right panel of Fig.\,\ref{fig:vsini_alpha}, we
plot measured magnetic flux $Bf$ as a function of $v\,\sin{i}$. If
magnetic flux did not saturate, we would expect the values of $Bf$ to
continue growing with higher rotation rate. This is not observed:
instead, the magnetic flux shows the same saturation effect as
H$\alpha$ emission.  From this result we conclude that magnetic flux
generation does not grow any further in stars with saturated H$\alpha$
emission. In other words, magnetic flux saturates in roughly the same
fashion as activity, implying that activity does not saturate solely
because the whole area of the star is covered with fields ($f=1$).

\subsection{Comparison to hotter stars}

\begin{deluxetable*}{llcccrrrrr}
  \tablecaption{\label{tab:Rossby} Masses, radii, projected rotational velocities, periods, and Rossby numbers, magnetic flux, and normalized X-ray luminosities for our sample stars and the stars from RB07 (see text).}
  \tablewidth{0pt}
  \tablehead{\colhead{Name} & \colhead{Other} & \colhead{SpT} & \colhead{$M/M_\sun$} & \colhead{$R/R_\sun$} & \colhead{$v\,\sin{i}$} & \colhead{$P/\sin{i}$} & \colhead{$\log{(Ro/\sin{i})}$} & \colhead{$Bf$ [G]}  &\colhead{$\log{\frac{L_{\rm X}}{L_{\rm bol}}}$} }
  \startdata

  GJ 3379       &          & M3.5 & 0.24 & 0.25 & $<$ 3 & $>$4.2\phantom{$^{a}$}  & $>-1.2$ &     2000 & $ -2.86$ \\
  GJ 2069\,B    & CV Cnc   & M4.0 & 0.24 & 0.25 &     6 &    2.3\phantom{$^{a}$}  & $ -1.5$ &     2500 & $ -2.77$ \\
  Gl 493.1      & FN Vir   & M4.5 & 0.17 & 0.20 &    18 &    0.5\phantom{$^{a}$}  & $ -2.1$ &     2100 & $ -3.31$ \\
  LHS 3376      &          & M4.5 & 0.14 & 0.17 &    16 &    0.5\phantom{$^{a}$}  & $ -2.1$ &     2000 & $ -3.63$ \\
  GJ 1154\,A    &          & M5.0 & 0.18 & 0.20 &     6 &    1.7\phantom{$^{a}$}  & $ -1.6$ &     2000 & $ -3.28$ \\
  GJ 1156       & GL Vir   & M5.0 & 0.14 & 0.16 &    17 &    0.5\phantom{$^{a}$}  & $ -2.1$ &     2100 & $ -3.39$ \\
  Gl 412\,B     & WX UMa   & M6.0 & 0.11 & 0.13 &     5 &    1.4\phantom{$^{a}$}  & $ -1.7$ & $>$ 3900 & $ -3.28$ \\
  \cutinhead{Stars from RB07}\\                                           
  Gl 70         &          & M2.0 & 0.35 & 0.33 & $<$ 3 & $>$5.6\phantom{$^{a}$}  & $>-1.1$ &        0 & $<-4.44$ \\
  Gl 729        & V1216 Sgr& M3.5 & 0.18 & 0.20 &     4 &    2.9\tablenotemark{a} & $ -1.4$ &     2200 & $ -3.50$ \\
  Gl 873        & EV Lac   & M3.5 & 0.33 & 0.31 &     3 &    4.4\tablenotemark{b} & $ -1.2$ &     3900 & $ -3.07$ \\
  Gl 388        & AD Leo   & M3.5 & 0.42 & 0.39 &     3 &    2.6\tablenotemark{b} & $ -1.4$ &     2900 & $ -3.02$ \\
  Gl 876        &          & M4.0 & 0.32 & 0.31 & $<$ 3 & $>$5.2\phantom{$^{a}$}  & $>-1.1$ &        0 & $ -5.23$ \\
  GJ 1005A      &          & M4.0 & 0.22 & 0.23 & $<$ 3 & $>$3.9\phantom{$^{a}$}  & $>-1.3$ &        0 & $ -5.05$ \\
  Gl 299        &          & M4.5 & 0.15 & 0.18 & $<$ 3 & $>$3.0\phantom{$^{a}$}  & $>-1.4$ &      500 & $<-5.55$ \\
  GJ 1227       &          & M4.5 & 0.17 & 0.19 & $<$ 3 & $>$3.2\phantom{$^{a}$}  & $>-1.3$ &        0 & $<-3.86$ \\
  GJ 1224       &          & M4.5 & 0.15 & 0.18 & $<$ 3 & $>$3.0\phantom{$^{a}$}  & $>-1.4$ &     2700 & $ -3.06$ \\
  Gl 285        & YZ CMi   & M4.5 & 0.31 & 0.30 &     5 &    2.8\tablenotemark{b} & $ -1.4$ & $>$ 3900 & $ -3.02$ \\
  Gl 905        & HH And   & M5.0 & 0.14 & 0.17 & $<$ 3 & $>$2.8\phantom{$^{a}$}  & $>-1.4$ &        0 & $ -3.75$ \\
  GJ 1057       &          & M5.0 & 0.16 & 0.18 & $<$ 3 & $>$3.1\phantom{$^{a}$}  & $>-1.4$ &        0 & $<-3.87$ \\
  GJ 1245B      &          & M5.5 & 0.11 & 0.14 &     7 &    1.0\phantom{$^{a}$}  & $ -1.8$ &     1700 & $ -3.58$ \\
  GJ 1286       &          & M5.5 & 0.12 & 0.14 & $<$ 3 & $>$2.4\phantom{$^{a}$}  & $>-1.5$ &      400 & $<-3.77$ \\
  GJ 1002       &          & M5.5 & 0.11 & 0.13 & $<$ 3 & $>$2.3\phantom{$^{a}$}  & $>-1.5$ &        0 & $<-5.24$ \\
  Gl 406        & CN Leo   & M5.5 & 0.10 & 0.13 &     3 &    1.9\phantom{$^{a}$}  & $ -1.6$ &     2400 & $ -2.77$ \\
  GJ 1111       & DX Cnc   & M6.0 & 0.10 & 0.12 &    13 &    0.4\phantom{$^{a}$}  & $ -2.2$ &     1700 & $ -3.88$ \\
  \enddata
\tablenotetext{a}{\cite{Kiraga07}}
\tablenotetext{b}{\cite{Saar01}}
\end{deluxetable*}

How does the saturation of magnetic flux in our M-dwarf sample fit
into the picture of rotation and activity in hotter stars? In order to
compare activity measurements in very different stars, the Rossby
number, i.e., the ratio of rotation period $P$ and convective overturn
time $\tau_{\rm conv}$, is the parameter of choice. We note, however,
that the main effect of the Rossby number is to compare rotation
periods, and that the influence of the convective overturn time is
still debatable. \citet{Kiraga07} provide a measurement of the
rotation period of Gl\,729. For Gl~873, AD~Leo, and YZ~CMi, $P$ can be
found in \citet{Saar01}. To obtain the (projected) Rossby number, $Ro
= P/\tau_{\rm conv}$, for the other stars, we calculate the
(projected) rotation periods for our sample and the stars from RB07
from the projected surface rotation velocity $v\,\sin{i}$ for which we
require the radii. Note that this only provides $Ro/\sin{i}$, which is
an upper limit of $Ro$. To determine the radius, we employed the
mass-luminosity relation from \citet{Delfosse00} and the mass-radius
relation at an age of 5\,Gyrs from \cite{Baraffe98}. To compute the
masses, we used J-magnitudes from \cite{2MASS} and distances from
\cite{Hawley96}.  Parallaxes for Gl~299 and GJ~1286 are from
\cite{Harrington80} and \cite{Oppenheimer01}, respectively. For the
convective overturn time, we adopt a value of $\tau_{\rm conv} =
70$\,d consistent with the values given for M dwarfs in
\citet{Saar01}, which are taken from \citet{Gilliland86}.
\citet{Kiraga07} have recently calculated empirical turnover times for
a sample of M stars. Their relation yields values about $\tau_c \sim
50$--100\,d for the mass range considered in our sample.  This is
roughly consistent with $\tau_c = 70$\,d; if we were using the
convective overturn times from \citet{Kiraga07}, the Rossby numbers
would be about 0.15\,dex larger (smaller) for the hottest (coolest)
stars. This difference does not affect our conclusions. Radii,
(projected) periods and Rossby numbers are given in
Table\,\ref{tab:Rossby}.

We combine our measurements with data on additional stars from
\citet{Saar96} and \citet{Saar01}. For the stars from \citet{Saar96}
we adopt a value of $\tau_{\rm conv} = 20$\,d (spectral type G0--K5)
noting that the convective overturn time changes over this range of
stars but that the difference is not significant on the level
considered here.

\begin{figure}
  \plotone{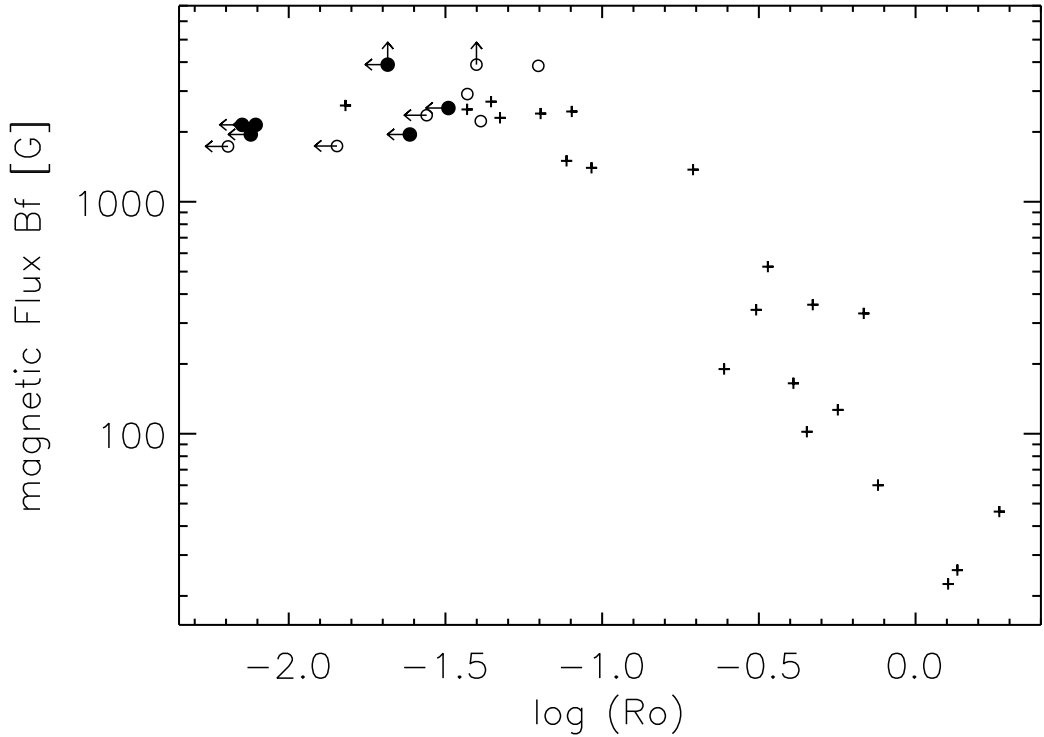}
  \caption{\label{fig:Bf_Ro}Magnetic flux $Bf$ as a function of Rossby
    number. Data are from \citet{Saar96}, \citet{Saar01}, RB07, and
    from this work. 11 stars without measured rotation periods and
    with no detection of rotational broadening ($v\,\sin{i} <
    3$\,km\,s$^{-1}$), i.e. lower limits of $Ro$, are not shown. They
    would form a vertical line at about $\log{Ro} = -1$ but probably
    lie on top of the rising part of the correlation
    \citep[see][]{Reiners07}. Data from this work are plotted as
    filled circles, data from RB07 as open circles (M stars), and data
    from \cite{Saar96} and \cite{Saar01} as crosses (spectral types
    G0--M2). Rotation rates from this work and from RB07 are
    calculated from $v\,\sin{i}$ implying that open and filled circles
    are $Ro/\sin{i}$ hence upper limits of $Ro$. }
\end{figure}

The behavior of $Bf$ with Rossby number is shown in
Fig.\,\ref{fig:Bf_Ro} where we plot all the mentioned samples
together. Data from this work are plotted as filled circles, data from
RB07 as open circles (M dwarfs), and data from \cite{Saar96} and
\cite{Saar01} as crosses (note that crosses are hotter stars). Stars
in which only upper limits of $v\,\sin{i}$ were measured are not shown
(non-detections of rotation).  Fig.\,\ref{fig:Bf_Ro} clearly shows the
dependence of magnetic flux generation with Rossby number: At large
Rossby numbers ($Ro > 0.1$), i.e. in the non-saturated regime,
magnetic flux increases with decreasing $Ro$ (more rapid rotation). At
smaller Rossby number, i.e., in the regime where H$\alpha$ and X-ray
emission saturate, $Bf$ saturates as well.

\begin{figure}
  \plotone{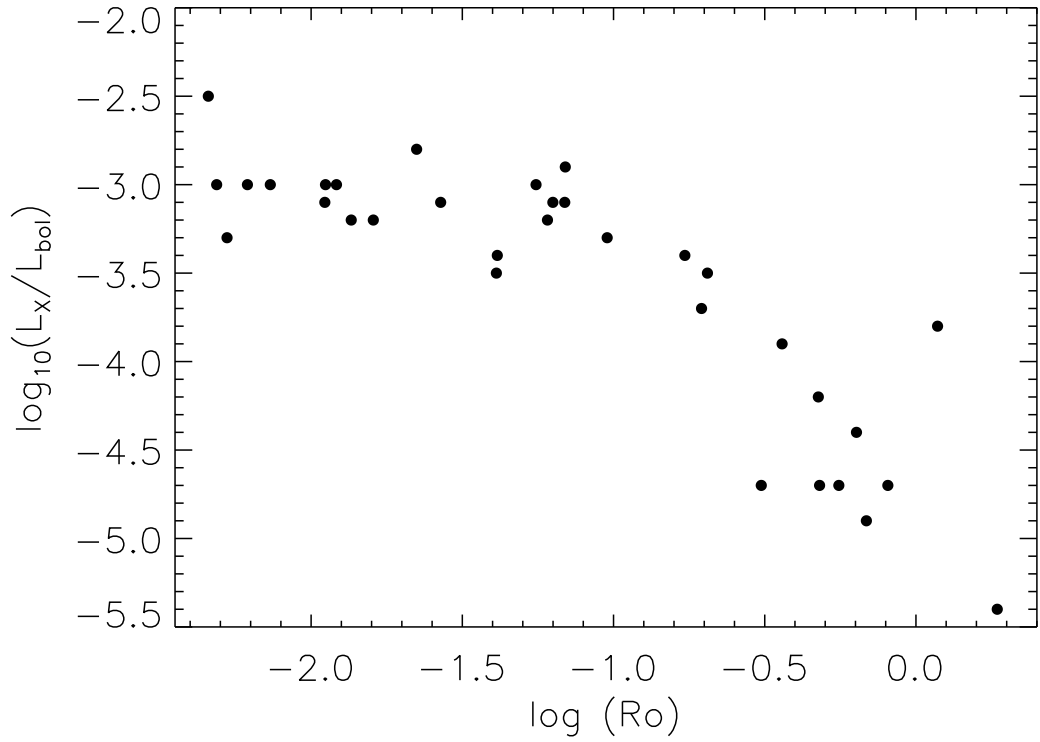}
  \caption{\label{fig:Kiraga}Normalized X-ray activity as a function
    of Rossby number in M stars (using $\tau_{\rm conv} = 70$\,d) from
    \citet{Kiraga07}. This plot is essentially the same as their
    Fig.\,7 but in logarithmic units so that the saturation plateau
    becomes clear.}
\end{figure}

In Fig.\,\ref{fig:Bf_Ro}, the saturated regime mainly consists of M
dwarfs while the rising part of the relation (crosses) is populated by
hotter stars. \citet{Reiners07} showed that a drop in activity with
higher Rossby number also appears in slowly rotating M dwarfs (his
Fig.\,10). A substantial piece of evidence for an intact
rotation-activity relation among M dwarfs can be found in the data
presented by \citet{Kiraga07}. From their data, we show normalized
X-ray activity vs. Rossby number (again assuming $\tau_{\rm conv} =
70$\,d) in Fig.\,\ref{fig:Kiraga}.  Clearly, a rise of activity at
high Rossby number ($Ro = 0.1\dots1$) and a saturation plateau at
lower Rossby number appears among M dwarfs, too.

We conclude that the dynamo process saturates at Rossby number of
about $Ro = 0.1$. No systematic increase of $Bf$ occurs if the Rossby
number is smaller by another order of magnitude.

\subsection{The regime of saturation}

The level of magnetic flux in the saturated regime is between 2 and
4\,kG. In two stars, we observe magnetic flux that may be higher than
4\,kG. At least one of the two stars is not among those with the
smallest Rossby numbers in our sample. We can speculate whether the
Rossby numbers of both stars with the highest magnetic flux are
smaller than the Rossby numbers of the other stars in our sample,
i.e., whether the Rossby numbers of YZ~CMi and WX~UMa are also on the
order of $\log{\emph{Ro}} \approx -2$. In the case of WX~Uma, we only
have $Ro/\sin{i}$, and for a real value of $\log{\emph{Ro}} = -2$ the
star would be observed under an inclination angle of $i < 30^\circ$.
For WX~UMa, this is a viable option. On the other hand, a rotation
period of $P = 2.8$\,d is reported for YZ~CMi, which is in good
agreement with an inclination angle close to $i = 90^\circ$ given the
estimated radius and the measured rotation velocity. To push the value
of $\log{\emph{Ro}} \approx -1.4$ to $-2$, either the rotation
velocity must be a factor of 4 higher ($i < 15^\circ$ implying that
the rotation period is wrong), or the convective overturn time must be
longer by the same amount. Both options seem rather unlikely. We note
that \citet{Saar01} reported a magnetic flux of $Bf = 3.3$\,kG for
YZ~CMi, a value that is somewhat lower than our result. This may
indicate that our value does not reflect an unusually strong average
field strength in YZ~CMi, but that the magnetic flux shows rather
large scatter (either due to uncertainties in the measurements or
temporal fluctuations).

The easiest explanation for the two very high values of $Bf > 4$\,kG
is that the scatter in the saturated magnetic flux level is fairly
large, and $Bf$ between 2 and 5\,kG might just be the allowed range at
small Rossby numbers (including observational effects). We have
searched for other parameters, such as exceptionally low gravity, that
could cause the high flux values in the two stars.  We did not find
any particular stellar parameter that distingiushes YZ~CMi and WX~Uma
from other flare stars. In particular, age is probably of little
direct importance for the generation of very high magnetic flux:
WX~Uma is an old disk flare star while YZ~CMi is a member of the young
disk population \citep{Veeder74}.

Even if the two strongest magnetic flux measurements are due to
exceptionally small Rossby numbers, this would not explain the
saturation of $Bf$ between $\log{\emph{Ro}} = -2$ and $-1$.  Across an
order of magnitude in $Ro$, $Bf$ varies by at most a factor of a few.
This is in striking contrast to the nearly hundredfold increase in
$Bf$ in going from $Ro \approx 1$ to $Ro \approx 0.1$.  We have
concluded that both magnetic flux and chromospheric H$\alpha$ (as well
as coronal X-ray) emission saturate at small $Ro$. With the large
uncertainties in $Bf$, we cannot reliably determine whether there is
any super-saturation effect, with magnetic flux declining in the most
rapid rotators.

\begin{figure}
  \plotone{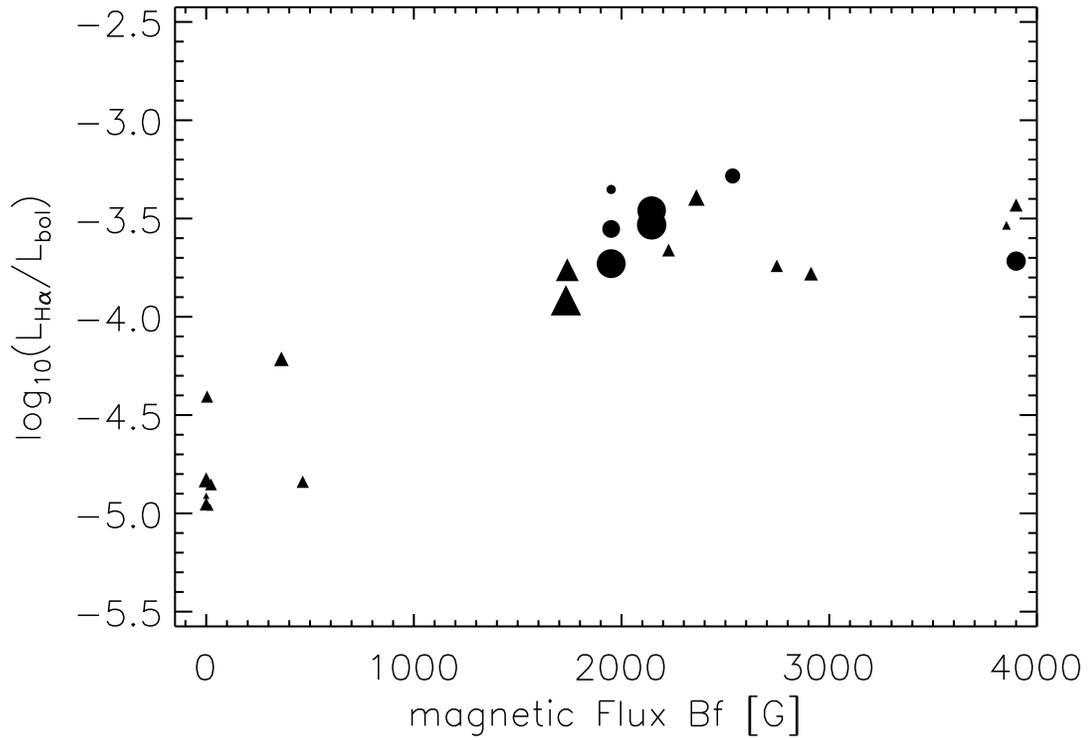}
  \caption{\label{fig:Bf_alpha}The correlation between H$\alpha$
    activity and magnetic flux $Bf$ in M stars. Symbol size scales
    with inverse Rossby number (large symbols have small $Ro$).
    Triangles are from RB07, circles are from this work.}
\end{figure}

Although both emission and magnetic flux appear to saturate at small
$Ro$, we observe a large scatter of magnetic flux among the stars with
small $Ro$, and likewise some scatter in H$\alpha$ emission. We thus
examined whether a relation between H$\alpha$ emission and magnetic
flux still exists at very small Rossby numbers. In
Fig.\,\ref{fig:Bf_alpha}, we plot normalized H$\alpha$ emission vs.
magnetic flux for our stars. As expected, $\log{L_{{\rm
      H}\alpha}/L_{\rm bol}}$ grows as $Bf$ grows from zero to 2\,kG.
Beyond that, i.e.  in the regime of saturated magnetic flux, no
further increase in chromospheric emission is observed, although $Bf$
grows as large as 4\,kG. No obvious correlation exists between
chromospheric emission and magnetic flux in the regime of saturation.
This could mean that saturation of chromospheric emission occurs above
a certain level of integrated flux on the star (for example, because
the filling factor reached unity). The field could then grow stronger
without causing further heating. This is in fact not ruled out by our
poorly constrained understanding of chromospheric heating mechanisms.

As of now, the available data do not permit us to distinguish reliably
between these two possibilities.  But, as noted above, the most rapid
rotators in our sample probably do not possess magnetic fluxes higher
than among other stars in the saturated regime: in
Fig.\,\ref{fig:Bf_alpha}, stars with the smallest Rossby numbers lie
at the low-$Bf$ end of the plateau.  Neither are they the stars with
the highest normalized chromspheric emission.  Thus the lack of tight
correlation between emission and $Bf$ in the saturation regime should
\emph{not} be construed as evidence that magnetic flux continues to be
correlated with rotation, but decouples from emission.  Instead, our
data suggest that any variation of H$\alpha$ emission or $Bf$ in the
saturation regime must either be random -- i.e. due to the
observational scatter -- or depend on parameters other than rotation.
At this stage, we note only that the plateau in
Fig.\,\ref{fig:Bf_alpha} is consistent with scatter alone.

\section{Summary}

Using absorption lines of molecular FeH we have measured the magnetic
flux in seven M stars that are known X-ray sources. With one exception
they were known as rapid rotators as well. We reinvestigated the
projected rotational velocities $v\,\sin{i}$ and measured
chromospheric emission in H$\alpha$. All stars proved to be strong
H$\alpha$ emitters. In our analysis of $v\,\sin{i}$, we found some
inconsistencies with former literature, which we ascribe to our more
sophisticated (direct) fitting method and better data quality.

All of our target stars show strong H$\alpha$ emission at the
saturation level ($\log{L_{{\rm H}\alpha}/L_{\rm bol}} > -4$). We
detected magnetic fields of kG-strength in all our targets. While in
less rapidly rotating stars, H$\alpha$ and magnetic flux correlate
with rotation rate, no such correlation is observed in our sample.

In contrast to sun-like stars of spectral types G and K, M dwarfs
rotating at $v\,\sin{i} \approx 10$\,km\,s$^{-1}$ have very small
Rossby numbers. Thus, our sample targets add to the amount of stars
with measured magnetic flux in the regime of small Rossby number, i.e.
in the regime of saturated magnetic activity.  Our primary goal was to
determine whether magnetic flux in this regime continues to grow with
more rapid rotation, as sometimes suggested \citep[e.g.,][]{Saar96},
or instead saturates in the same manner as coronal and chromospheric
emission.

Our main conclusion is that around a Rossby number of $Ro \sim 0.1$,
magnetic flux saturates at approximately $Bf = 3$\,kG. Below $Ro \sim
0.1$, $Bf$ does not grow stronger with decreasing Rossby number.  In
looking at the effect of Rossby number ($Ro = P/\tau_{\rm conv}$), we
are primarily sensitive to the effects of rotation, because the
sampled rotations vary by a factor of more than ten, while the
convective overturn time probably changes less than a factor of two in
the range of M stars we are considering.

In the regime of saturated magnetic flux and chromospheric emission,
we still observe a strong scatter in magnetic flux. The interpretation
of this feature is not clear. It may be that below the critical Rossby
number, normalized H$\alpha$ emission is not sensitive to changes in
$Bf$. Alternatively, in light of the large systematic errors of our
$Bf$ measurements, the scatter of $Bf$ in the saturated regime may be
fully explained by observational uncertainties.

Our results indicate that the strengths of stellar magnetic fields,
and not merely their filling fractions on the surface alone, reach a
maximum at a certain rotation rate. The scatter in magnetic flux among
the stars with very small Rossby numbers, however, is substantial.
Typical values are between 2 and 4\,kG. Two stars show magnetic flux
stronger than 4\,kG, and we cannot exclude that individual stars can
generate magnetic flux stronger than that. Determining what sets the
maximum value -- whether equipartition with either the atmospheric
pressure or the turbulent velocity field, or some more subtle effect
-- remains a challenge for future observations and theories.

\acknowledgments 

This work is based on observations from the W.M. Keck Observatory,
which is operated as a scientific partnership among the California
Institute of Technology, the University of California and the National
Aeronautics and Space Administration. We would like to acknowledge the
great cultural significance of Mauna Kea for native Hawaiians and
express our gratitude for permission to observe from atop this
mountain.  AR acknowledges research funding from the DFG under an Emmy
Noether Fellowship (RE 1664/4-1). G.B. acknowledges support from the
NSF through grant AST-0606748. M.B. was supported by an NSF Astronomy
and Astrophysics postdoctoral fellowship (AST-0502413).







\end{document}